\documentclass[a4paper,fleqn,usenatbib]{mnras}

\usepackage[T1]{fontenc}
\usepackage{ae,aecompl}

\usepackage{graphicx}	
\usepackage{amsmath}	
\usepackage{amssymb}	

\title[O/H of Neptune]{What is Neptune's D/H ratio really telling us about its water abundance ?}

\author[M. Ali-Dib \& G. Lakhlani]{
Mohamad Ali-Dib$^{1}$$^{,}$$^{2}$\thanks{E-mail: m.alidib@utoronto.ca}
and Gunjan Lakhlani$^{2}$$^{,}$$^{3}$
\\
$^{1}$Centre for Planetary Sciences, Department of Physical \& Environmental Sciences, University of Toronto at Scarborough,\\
Toronto, ON M1C 1A4, Canada\\
$^{2}$Canadian Institute for Theoretical Astrophysics, 60 St. George St, University of Toronto, Toronto, ON M5S 3H8, Canada\\
$^{3}$Department of Physics, University of Toronto, Toronto, ON M5S 1A7, Canada
}

\date{Accepted XXX. Received YYY; in original form ZZZ}

\pubyear{2016}

\begin{document}
\label{firstpage}
\pagerange{\pageref{firstpage}--\pageref{lastpage}}
\maketitle

\begin{abstract}
We investigate the deep water abundance of Neptune using a simple 2-{component} (core + envelope) toy model. The free parameters of the model are the total mass of heavy elements in the planet ($Z$), the mass fraction of $Z$ in the envelope ($f_{env}$), and the D/H ratio of the accreted building blocks ($D/H_{\text{build}}$). We systematically search the allowed parameter space on a grid and constrain it using Neptune's bulk carbon abundance, D/H ratio, and interior structure models. Assuming solar C/O ratio and cometary D/H for the accreted building blocks forming the planet, we can fit all of the constraints if less than $\sim$ 15\% of $Z$ is in the envelope ($f_{env}^{median} \sim$ 7\%), and the rest is locked in a solid core. This model predicts a maximum bulk oxygen abundance in Neptune of 65$\times$ solar value. If we assume a C/O of 0.17, corresponding to clathrate-hydrates building blocks, we predict a maximum oxygen abundance of 200$\times$ solar value with a median value of $\sim 140$. Thus, both cases lead to an oxygen abundance significantly lower than the preferred value of \cite{cavalie} ($\sim$ 540 $\times$ solar), inferred from model dependent deep CO observations. Such high water abundances are excluded by our simple but robust model. We attribute this discrepancy to our imperfect understanding of either the interior structure of Neptune or the chemistry of the primordial protosolar nebula.

\end{abstract}



\section{Introduction}
Uranus and Neptune are arguably the least understood planets in the solar system. Their formation mechanism \citep{robinson,helled,ali-diba}, evolution, and current states are poorly understood due to the sparsity of available data. Among the few chemical elements with constrained abundances in the icy giants' atmopsheres are carbon (through methane observations \cite{baines}), D/H ratio (measured by \textit{Herschel} \citep{feucht}), and finally CO \citep{lellouch2005}. However, these observations can go a long way in revealing the histories of these planets. Since methane has also been constrained in Jupiter and Saturn \citep{mousisprobe}, its abundance in different planets can be compared and used as a robust tracer of atmospheric metallicity. The D/H ratio on the other hand has a long history of being used as a tracer for the formation temperature of ices \citep{mousis0,ali-dibch}. This is because water vapors that went through high temperatures phases in the presence of the disk's H$_2$ before re-condensing into ices should have lower D/H ratio than those which did not undergo this heating, due to the chemical reaction:
\begin{equation}
\text{HDO} + \text{H$_2$} \leftrightarrows \text{H$_2$O} + \text{HD}
\end{equation}
that favor HDO's transformation into H$_2$O at high temperatures.

CO is a trace non equilibrium specie that can form starting from water and methane at high temperatures in the inner envelope, and quickly dissolve back into these elements when transported to the colder outer parts of the atmosphere:
\begin{equation}
\label{CO}
\text{CO} + \text{3H$_2$} \leftrightarrows \text{CH$_4$} + \text{H$_2$O}
\end{equation}

The detection of CO in the ice giants' outer atmospheres hence implies significant vertical mixing between the inner and outer parts of the envelope, and the presence of important amounts of water in the deep interior of the planets. In theory, CO tropospheric abundance can therefore be used to constrain the bulk oxygen abundance, that is not accessible through direct observations due to the cold trap at the tropopause level. {In reality, this is however complicated since both a complex chemical model to the multi-stages reaction \ref{CO} and assumptions on the efficiency of the atmospheric vertical mixing (K$_{zz}$) are needed in order to connect the measured CO abundance to the deep water content of the planet. CO can be observed only if the chemical timescale for its destruction through equation \ref{CO} is longer than the vertical mixing timescale.} This technique was initially used by \citep{loddersfeg} to constrain the bulk water abundance in Neptune to $\sim$ 400$\times$ solar value. More recently, \cite{cavalie} used IRAM-30m and \textit{Herschel/SPIRE} observations, with an updated chemical and vertical mixing scheme to constrain Neptune's deep water abundance to a best value of $\sim$ 540 $\times$ solar. They however note the difficulty in reconciling this value with formation and interior structure models.

In this work we use a simple 2-{component} (solid core + envelope) toy model to fit these observational constraints. We focus exclusively on Neptune, since we only have upper limits on the CO and water abundances in Uranus. The main question we try to answer is \textit{What bulk oxygen abundance values are compatible simultaneously with the carbon abundance, D/H ratio, and interior structure models of Neptune ?}

\section{The toy model}

We use a simple toy model that nonetheless captures the main processes at play. We start with a 2-{component} planet with a solid core and an envelope. It has a total mass fraction of heavy elements (including those locked up in the core in addition to the atmospheric metals) $Z$ and we assume that the solid building blocks that formed the planet have a given D/H ratio value of $D/H_{\text{build}}$. We assume that $Z = Z_{\text{env}} + Z_{\text{core}}$ is distributed between the solid core and the envelope, where the atmospheric heavy elements mass fraction is $Z_{\text{env}}$ and we define $f_{\text{env}} = Z_{\text{env}} / (Z_{\text{env}} + Z_{\text{core}})$. $f_{\text{env}}$ takes values ranging from 0 if the heavy elements are locked entirely in the solid core, to 1 if there is no solid core at all (planet completely mixed). $f_{\text{env}}$ hence depends on the formation and evolution of the planet, where processes like core erosion and atmospheric enrichment through the dissolution of pebbles during accretion can increase it. Our toy model hence has 3 free parameters: $Z$,  $f_{\text{env}}$, and  $D/H_{\text{build}}$. We vary these parameters on a grid as shown in table \ref{t1}, and use them to calculate the final atmospheric D/H for the planet:
\begin{equation}
D/H_{\text{p}}=(1-x_{\text{H}_\text{2}})\times D/H_{\text{build}} +(x_{\text{H}_\text{2}} + x_{\text{CO}})\times D/H_{\text{H}_\text{2}}
\end{equation}
{where $D/H_{\text{H}_\text{2}}=2.25 \times 10^{-5}$ is the protoplanetary disk's HD/H$_2$ ratio as measured in Jupiter and Saturn's atmospheres \citep{lellouch}, and $x_{H_2}$ and $x_{CO}$ are respectively the volumetric (molar number) ratios of $H_2$ and CO defined as:}
\begin{equation}
x_{\text{H}_\text{2}} + x_{\text{CO}} = \frac{ \frac{m_{CO}}{M_{CO}} + \frac{m_{H_2}}{M_{H_2}} } { \frac{m_{CO}}{M_{CO}} +\frac{m_{H_2}}{M_{H_2}} +\frac{m_{He}}{M_{He}} + \frac{m_{H_{2O}}}{M_{H_2O}} }
\end{equation}
{with $M_{X}$ is the molar mass of $X$} 
{and $m_{X}$ is the total mass of $X$ in the envelope at formation time (so CO in this equation would be the primordially accreted CO, not the non-equilibrium trace CO currently present in Neptune's atmosphere). Note that CO molecule transforming into water following reaction \ref{CO} will have the disk's gas D/H value ($D/H_{\text{H}_\text{2}}$).}  
{We start from $m_{water} + m_{CO} =  I_f \times Z_{env} \times m_{Neptune}$ and $C/O = n_{CO} / (n_{H_2O} + n_{CO})$ where $C/O$ is the carbon to oxygen molar ratio, n$_{X}$ = m$_{X}$/M$_{X}$, and $I_f$ = $m_{ices, total}^{env} / (m_{ices, total}^{env} + m_{rock}^{env})$ is the envelope's ice mass fraction. Hence $I_f$ = 0.5 in most comets where $m_{ices, total} \sim m_{rock}$ \citep{mumma}. In this case we are implicitly assuming that only ices are contributing to the atmospheric chemistry, i.e. rocks did not melt during Neptune's formation. Whether these rocks remain suspended in the atmosphere or settle to the core will not affect the gas phase chemistry we are interpreting. We also try a case with $I_f$ = 1, hence assuming that both ices and rocks sublimated during formation and hence both contribute to the atmospheric chemistry.\footnote{{Technically $I_f$ = 1 implies that Neptune formed entirely from ices with contribution from rocks, however since we assume similar chemical composition between these two components, setting $I_f$ to 1 will have the same effect as assuming that both rocks and ices are vaporized during accretion}.} For simplicity we are assume that the C/O ratio is similar between the ices and rocks phases (C/O$_{ice}$ = C/O$_{rock}$ = C/O).}
{From these two equations we calculate the total water mass as:}
\begin{equation}
m_{water} = \frac{1}{1+ \frac{1.55 \times (C/O)}{(1 - (C/O))}} \times I_f \times Z_{env} \times m_{Neptune}
\end{equation}
{and the total CO mass as:}
\begin{equation}
m_{CO} = I_f \times Z_{env} \times m_{Neptune} - m_{water}
\end{equation}
{Finally, we calculate the bulk oxygen and carbon molar abundances with respect to hydrogen (and normalized to the solar values \citep{asplund}) as:}
\begin{equation}
X_{Oxygen}/X_{Oxygen}^{\odot} = \bigg( \frac{m_{water}}{M_{water}} + \frac{ m_{CO}}{M_{CO}} \bigg)\times \frac{M_{H_2}}{m_{H_2}}
\end{equation}
and:
\begin{equation}
X_{Carbon} / X_{Carbon}^{\odot} = \bigg( \frac{ m_{CO}}{M_{CO}} \bigg)\times \frac{M_{H_2}}{m_{H_2}}
\end{equation}

\section{Results \& Discussions}
\subsection{Case: Nominal}
\begin{figure*}
\begin{centering}
	\includegraphics[scale=0.30]{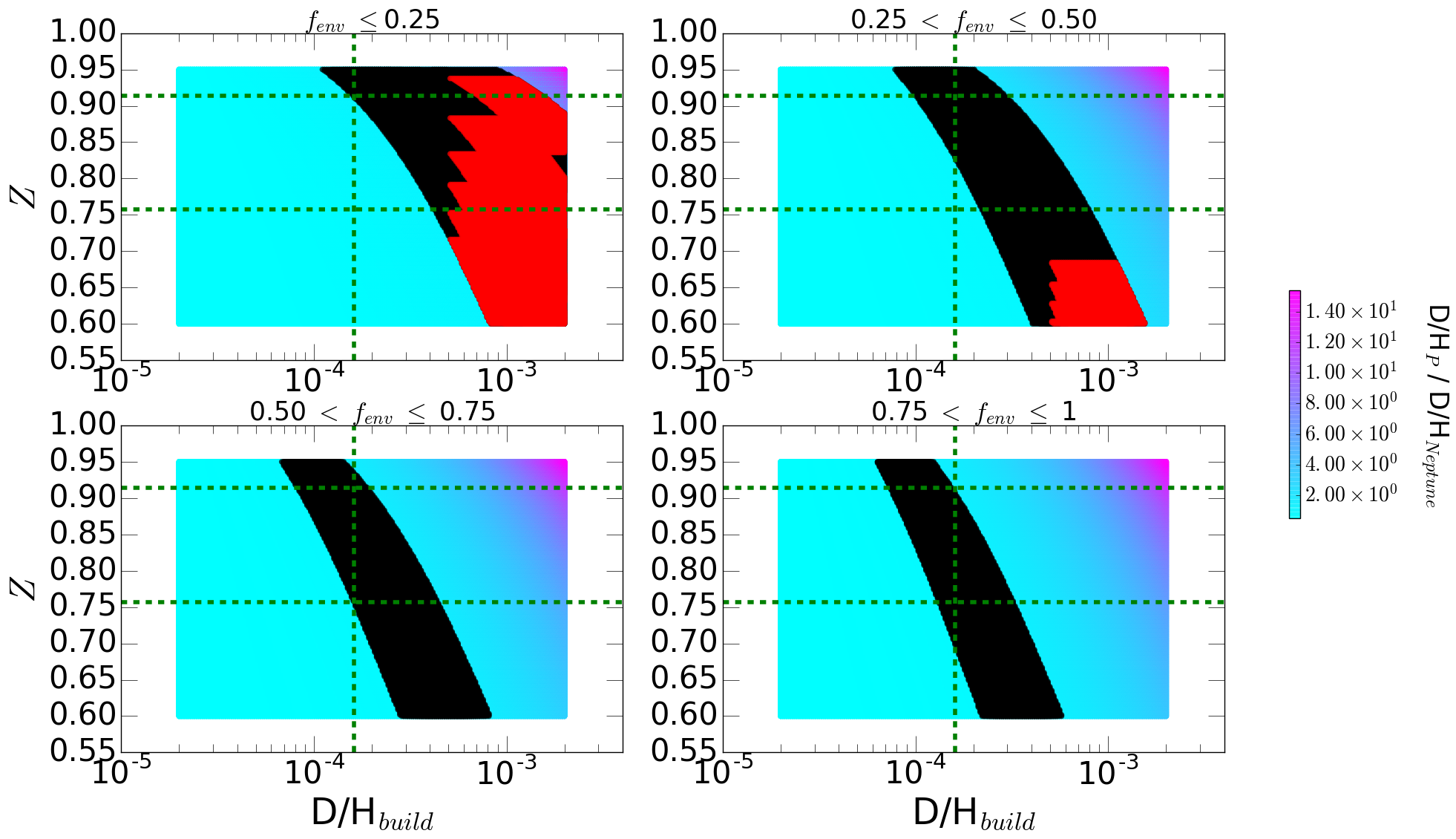}
   \caption{Planets final atmospheric D/H ratio as a function of the toy model's free parameters: the total mass of heavy elements (core + envelope), the D/H of the building blocks, and the fraction of heavy elements in the envelope (binned into 4 categories). The two horizontal dashed lines are the lower and upper limit on $Z$ from \citep{helled1}. The vertical dashed line is the VSMOW D/H ratio. In black are regions of the parameter space consistent with Neptune's measured D/H value. In red are regions consistent with Neptune's carbon abundance. This is the case with C/O = 0.55 (solar value).}
    \label{fig:main}
    \end{centering}
\end{figure*}

\begin{figure}
	\begin{centering}
		\includegraphics[scale=0.45]{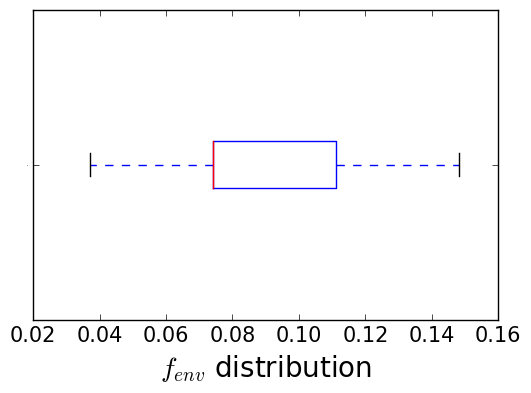}
		\caption{The distribution of the fraction of heavy elements in the envelope $f_{\text{env}}$ for planets fitting all observational constraints. }
		\label{fig:fenv}
	\end{centering}
\end{figure}

\begin{figure}
	\begin{centering}
		\includegraphics[scale=0.45]{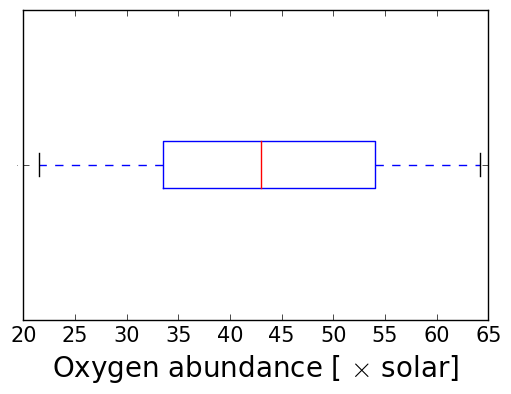}
		\caption{The distribution of the predicted oxygen abundance in Neptune for models consistent with all of the measurements. This plot puts an upper limit of 65 $\times$ solar value on the oxygen abundance. }
		\label{fig:second}
	\end{centering}
\end{figure}

Results are presented in Fig. \ref{fig:main} showing the atmospheric D/H of a planet ($D/H_{\text{p}}$) as a function of the toy model's free parameters. {This is for our nominal case with solar C/O = 0.55 and $I_f$ = 1 (thus assuming that both rocks and ices vaporize during accretion).}
$D/H_{\text{p}}$ increases with higher values of $Z$ or $f_{\text{env}}$ since this implies depositing more (D-rich) solids into the envelope. It also increases with $D/H_{\text{build}}$, since this implies a higher Deuterium abundance in the deposited solids. The plot shows that we can fit Neptune's atmospheric D/H ratio for a wide region in parameter space (highlighted in black). The other observational constraints however shrink the allowed space. We can first exclude models with $D/H_{\text{build}}$ lower than Earth's VSMOW value (vertical dashed line). This is the same value found in most CI chondrites (associated to the relatively volatiles rich C-type asteroids), and is simultaneously the lowest value measured in a comet (103P/Hartley measured with \textit{Herschel}) \citep{robert,ali-dibch,altwegg}. Hence, it is very unlikely that Neptune formed from building blocks with $D/H_{\text{build}}$ lower than this value, even if these blocks were rock-dominated. 
The second constraint to add is the bulk carbon abundance of Neptune. We compare the resulting C/H ratio of the planet to the value measured in Neptune (20 to 60 $\times$ solar \citep{baines,cavalie}). Compatible models are shown in red on the plot. While this constraint on its own is not very restrictive, it can be used along with the constraints we have on $Z$ to significantly shrink the allowed parameter space. Regions in Fig. \ref{fig:main} not boxed by the horizontal dashed lines have their total mass of heavy elements $Z$ is outside the range consistent with $J_2$ and $J_4$ measurements and other \textit{Voyager} data as modeled by \cite{helled1}. For these limits, we chose the lowest and highest $Z$ values found by \cite{helled1}, including all their consistent models. The lower limit is thus $Z$ found through their ``case I'' (metallicity linearly increasing toward the center) for a fully SiO$_2$ interior, and the upper limit is the value in their ``case II'' (``classic 3-{layer} interior) assuming a fully H$_2$O interior. These two constraints (C/H and $Z$) together exclude models with $f_{\text{env}}$ higher than 25\%. The only parameter space region remaining is thus the red part sandwiched between the 2 horizontal lines in the top left panel of the plot. \\
We now focus exclusively on planets compatible with all of our constraints. First we notice that many of these planets have $Z$ values below 0.88, the minimal value found by \cite{helled1} for a pure water ices interior. Moreover, the planets with $Z$ higher than 0.88 all have very low $f_{\text{env}} \sim$ 0.07, which seem unlikely but cannot be fully excluded. This implies that a significant fraction of Neptune's interior should be rocky, which is expected since comets are on average $\sim$ 50\% rocky by mass \citep{jess}. In Fig. \ref{fig:fenv} we plot the distribution of $f_{\text{env}}$ for planets fitting the constraints, and calculate a low median value of $\sim$ 7\%. Interior structure models have long predicted that convection is inefficient in Neptune-mass planets (\cite{podolak1995, guillot1994},\cite{wilsonmilitzer,vazan}), what might contribute to these low values of $f_{\text{env}}$. However, for $f_{\text{env}}$ to be this low, accreted solids should additionally not dissolve in the envelope during accretion, what seem at odd with formation models \citep{podolak}. In Fig. \ref{fig:second} we show the distribution of the predicted water abundance of these planets. The maximal value found by our toy model is around 65 $\times$ solar, a factor 8 lower than the best value of \cite{cavalie}. This is however expected since we initially assumed a solar C/O ratio of 0.55 for the accreted solids. Now we relax this assumption to test its effects on the results.

\subsection{Case: Clathrates}
\begin{figure}
\begin{centering}
	\includegraphics[scale=0.45]{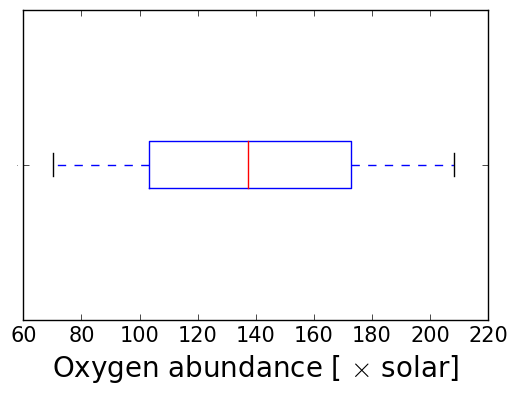}
   \caption{Same as Fig. \ref{fig:second}, but with building blocks C/O = 0.17, corresponding to the clathrates case.}
    \label{fig:clathrate}
    \end{centering}
\end{figure}

We ran the same calculations as above but assuming that Neptune's building blocks were entirely water clathrates, trapping the other volatiles (in this case CO) \citep{lunine, mousis1,mousis2,mousis3}. In clathrates theory, 5.75 atoms of water are needed to create the cage trapping CO, leading to a C/O ratio $\sim 0.17$ for the accreted building blocks. Oxygen abundances for planets fitting all other criteria (D/H, C/H and $Z$) for this case  are shown in Fig. \ref{fig:clathrate}. We notice that even in this extreme case, the oxygen abundance is never higher than $\sim$ 200 $\times$ solar, 2.5 times less than the best value of \cite{cavalie}, with the median value being around 130. The median $f_{\text{env}}$ value for this case is around 20\%, higher than the case with solar C/O. This is because more solids are needed to fit the carbon abundance due to the low C/O ratio. \\
These low values for $f_{\text{env}}$ are in contrast with \cite{ali-diba} who assumed fully mixed planets ($f_{\text{env}}$=1). Moreover, \cite{ali-diba} predicted C/O=1 for both Uranus and Neptune, while this model assumed C/O of 0.5 and 0.17 a priori. \cite{cavalie} on the other hand found C/O$\sim 0.03$. A robust, model independent, measurement of C/O and solid core mass in Neptune through a dedicated mission are hence necessary to distinguish between these models.

\subsection{Case: No rocks contribution }

\begin{figure*}
\begin{centering}
	\includegraphics[scale=0.30]{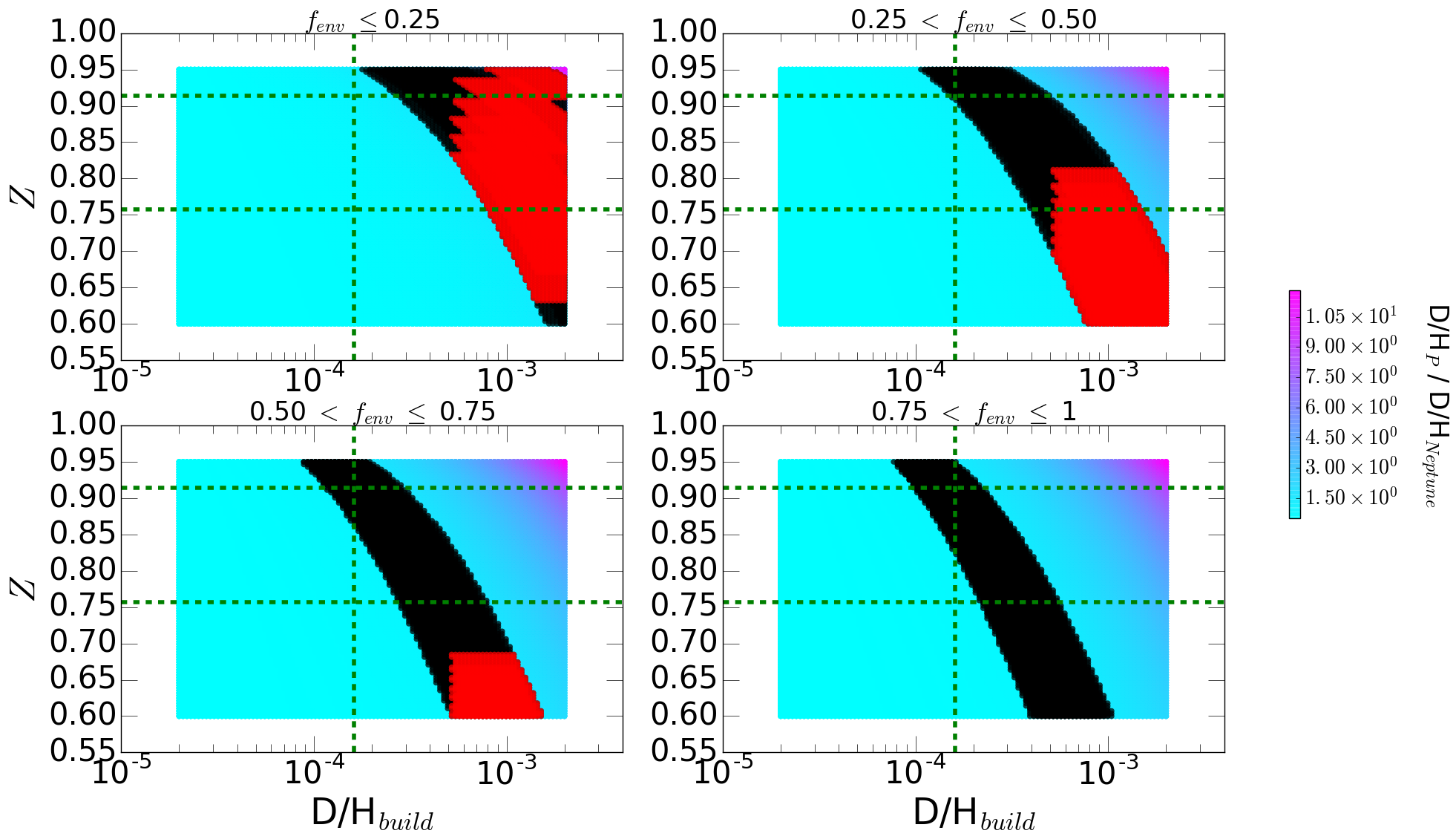}
   \caption{Same as Fig. \ref{fig:main}, but with $I_f = 0.5$, hence assuming that only ices will contribute to the atmospheric chemistry. Since the total amount of solids contributing to the chemistry is now less by half, higher $E_f$ values are allowed. }
    \label{fig:main2}
    \end{centering}
\end{figure*}

{The last parameter we vary is $I_f$, that we set initially to 1. We now set $I_f = 0.5$ hence assuming that only ices will contribute to the atmospheric chemistry of Neptune, that is rocks will not vaporize during accretion. These rocks can settle to the core later during the planet's evolution, but even if they are small enough to stay coupled to the gas in the atmosphere, they will not contribute to its chemistry unless they vaporize which is unlikely to happen under the pressure-temperature conditions of present day Neptune. Results for this case are shown in Fig. \ref{fig:main2}. The main difference here is that since the amount of solid material contributing to the atmospheric chemistry is only half of that of our nominal case (since rocks are no longer contributing), the mass fraction of metals in the envelope $f_{\text{env}}$ can be significantly higher than for the nominal case (up to $\sim$ 33\% with a median of 15\%). This is because a higher fraction of the available solids need to contribute to the atmospheric chemistry if the total amount of available materials is less. This case however leads to the same oxygen abundance distribution with a median of $\sim 45 \times$solar as our nominal case. The same trends are found in the case with C/O = 0.17 and $I_f = 0.5$. }

\subsection{Sanity check}
{Since, for all of three cases, no region of the allowed parameter space predicted an oxygen abundance consistent with the findings of \cite{cavalie}, we reverse the problem as a sanity check. We therefore assume a priori an oxygen abundance of 540 $\times$ solar, and thus calculate a corresponding $Z$ value of 0.83. 
This is however the amount of solids needed in the envelope to increase its water abundance to 540 $\times$ solar, and hence correspond to the case with $f_{\text{env}}$ = 1 (completely well mixed planet with no central core). 
However, by looking at Fig. \ref{fig:main}, we find no model in the bottom right panel (where $f_{\text{env}}$ = 1) that fits the carbon abundance of Neptune for this $Z$ value. We make same conclusions as above for the clathrates case. Hence, no region of parameter space in our toy model leads to such extremely high oxygen abundance.} \\
{This discrepancy can be possibly solved if Neptune's interior structure is significantly different than the models used by \cite{cavalie} and/or \cite{helled1}. This will lead to a different thermal structure and $K_{zz}$ values, possibly changing the retrieved oxygen abundance. Another caveat is the chemical assumptions of this model where we assumed that water and CO are the only ices present, and that ices and rocks have similar C/O ratios. Finally, as \cite{cavalie} and \cite{wang} mentioned, when constraining the oxygen abundance from CO observations, different chemical network assumptions will lead to significantly different results. }
\section{Summary \& Conclusions}

We used a simple 2-{component} (core + envelope) toy model for Neptune to predict its bulk oxygen abundance starting from its measured chemical composition (D/H and C/H ratios), and interior structure models. We then compared it to the recently published values found by \cite{cavalie} from modeling tropospheric CO observations. 

The model's free parameters are the total mass of heavy elements in the planet $Z$ that we allow to vary from 0.75 to 0.92 as constrained by \cite{helled1}, the mass fraction of $Z$ in the envelope ($f_{env}$) that we vary between 0 and 1, and the D/H ratio of the accreted building blocks that we vary across the entire range found in comets. We finally try two values for C/O: 0.55 corresponding to the solar case, and 0.17 corresponding to a case where all building blocks are clathrates.

For the solar case, we find a maximal allowed bulk oxygen abundance in Neptune of 65$\times$ solar value, while in the clathrates case this can be as high as 200$\times$ solar value. Both cases hence give oxygen abundance significantly lower than the best value found by \cite{cavalie} of $\sim 540$. Moreover, both cases predict a massive solid core decoupled from envelope, with $f_{env} <$ 25\%. Exoplanets observations have showed that Super Earths to Neptune-mass planets are the most common in the galaxy. A dedicated Uranus \& Neptune mission \citep{np1,np2} will hence shed light on not only the history of our solar system, but also the formation of exoplanets.

\renewcommand\arraystretch{1.2}
\begin{table}
\begin{center}
\caption{Free parameter space.}
\footnotesize
{\begin{tabular}{lcccc}
\hline
\noalign{\smallskip}
Parameter			& Range			& Step	\\
\hline
 $f_{\text{env}}$			& 0 - 1		& 0.03 	 \\			

$Z$ & 0.6 - 0.95 &	 0.0017 \\

$D/H_{\text{build}}$			& 2$\times 10^{-5}$	- 2$\times 10^{-3}$		& 200 log uniform points		\\	

\hline	
\end{tabular}}\\

\label{t1}
\end{center}
\end{table}

\section*{Acknowledgements}
We thank R. Helled for useful discussions on the interior structure of Neptune. We thank the anonymous referee for useful comments that improved this manuscript.





\begin{thebibliography}{00}





\bibitem[\protect\citeauthoryear{Ali-Dib et al.}{2014a}]{ali-diba} Ali-Dib M., Mousis O., Petit J.-M., Lunine J.~I., 2014, ApJ, 793, 9 


\bibitem[\protect\citeauthoryear{Ali-Dib et al.}{2015}]{ali-dibch} Ali-Dib M., Martin R.~G., Petit J.-M., Mousis O., Vernazza P., Lunine J.~I., 2015, A\&A, 583, A58 



\bibitem[\protect\citeauthoryear{Altwegg et al.}{2015}]{altwegg} Altwegg K., et al., 2015, Sci, 347, 1261952 

\bibitem[\protect\citeauthoryear{Arridge et al.}{2012}]{np1} Arridge C.~S., et al., 2012, ExA, 33, 753 

\bibitem[\protect\citeauthoryear{Arridge et al.}{2014}]{np2} Arridge C.~S., et al., 2014, P\&SS, 104, 122 



\bibitem[\protect\citeauthoryear{Asplund et al.}{2009}]{asplund} Asplund M., Grevesse N., Sauval A.~J., Scott P., 2009, ARA\&A, 47, 481 



\bibitem[\protect\citeauthoryear{Baines et al.}{1995}]{baines} Baines K.~H., Mickelson M.~E., Larson L.~E., Ferguson D.~W., 1995, Icar, 114, 328 










\bibitem[\protect\citeauthoryear{Cavali{\'e} et al.}{2017}]{cavalie} Cavali{\'e} T., Venot O., Selsis F., Hersant F., Hartogh P., Leconte J., 2017, Icar, 291, 1 

\bibitem[\protect\citeauthoryear{Dodson-Robinson \& Bodenheimer}{2010}]{robinson} Dodson-Robinson S.~E., Bodenheimer P., 2010, Icar, 207, 491 

\bibitem[\protect\citeauthoryear{Feuchtgruber et al.}{2013}]{feucht} Feuchtgruber H., et al., 2013, A\&A, 551, A126 


\bibitem[\protect\citeauthoryear{Guillot et al.}{1994}]{guillot1994} Guillot T., Gautier D., Chabrier G., Mosser B., 1994, Icar, 112, 337 


\bibitem[\protect\citeauthoryear{Helled et al.}{2011}]{helled1} Helled R., Anderson J.~D., Podolak M., Schubert G., 2011, ApJ, 726, 15 

\bibitem[\protect\citeauthoryear{Helled \& Bodenheimer}{2014}]{helled} Helled R., Bodenheimer P., 2014, ApJ, 789, 69 
\bibitem[\protect\citeauthoryear{Jessberger, Christoforidis, \& Kissel}{1988}]{jess} Jessberger E.~K., Christoforidis A., Kissel J., 1988, Natur, 332, 691 






\bibitem[\protect\citeauthoryear{Lellouch et al.}{2001}]{lellouch} Lellouch E., B{\'e}zard B., Fouchet T., Feuchtgruber H., Encrenaz T., de Graauw T., 2001, A\&A, 370, 610 

\bibitem[\protect\citeauthoryear{Lellouch, Moreno, \& Paubert}{2005}]{lellouch2005} Lellouch E., Moreno R., Paubert G., 2005, A\&A, 430, L37 


\bibitem[\protect\citeauthoryear{Lodders \& Fegley}{1994}]{loddersfeg} Lodders K., Fegley B., Jr., 1994, Icar, 112, 368 

\bibitem[Lunine \& Stevenson(1985)]{lunine} Lunine, J.~I., \& Stevenson, D.~J.\ 1985, \apjs, 58, 493 
\bibitem[\protect\citeauthoryear{Mumma \& Charnley}{2011}]{mumma} Mumma M.~J., Charnley S.~B., 2011, ARA\&A, 49, 471 

\bibitem[\protect\citeauthoryear{Mousis et al.}{2000}]{mousis0} Mousis O., Gautier D., Bockel{\'e}e-Morvan D., Robert F., Dubrulle B., Drouart A., 2000, Icar, 148, 513 


\bibitem[\protect\citeauthoryear{Mousis et al.}{2010}]{mousis1} Mousis O., Lunine J.~I., Picaud S., Cordier D., 2010, FaDi, 147, 509 

\bibitem[\protect\citeauthoryear{Mousis et al.}{2014}]{mousis2} Mousis O., Lunine J.~I., Fletcher L.~N., Mandt K.~E., Ali-Dib M., Gautier D., Atreya S., 2014, ApJ, 796, L28 
\bibitem[\protect\citeauthoryear{Mousis et al.}{2014}]{mousisprobe} Mousis O., et al., 2014, P\&SS, 104, 29 


\bibitem[\protect\citeauthoryear{Mousis et al.}{2016}]{mousis3} Mousis O., et al., 2016, ApJ, 819, L33 




\bibitem[\protect\citeauthoryear{Podolak, Pollack, \& Reynolds}{1988}]{podolak} Podolak M., Pollack J.~B., Reynolds R.~T., 1988, Icar, 73, 163 

\bibitem[\protect\citeauthoryear{Podolak, Weizman, \& Marley}{1995}]{podolak1995} Podolak M., Weizman A., Marley M., 1995, P\&SS, 43, 1517 

\bibitem[\protect\citeauthoryear{Robert}{2006}]{robert} Robert F., 2006, mess.book, 341 

\bibitem[\protect\citeauthoryear{Vazan et al.}{2016}]{vazan} Vazan A., Helled R., Podolak M., Kovetz A., 2016, ApJ, 829, 118 



\bibitem[\protect\citeauthoryear{Wang et al.}{2015}]{wang} Wang D., Gierasch P.~J., Lunine J.~I., Mousis O., 2015, Icar, 250, 154 


\bibitem[\protect\citeauthoryear{Wilson \& Militzer}{2012}]{wilsonmilitzer} Wilson H.~F., Militzer B., 2012, ApJ, 745, 54 





\end{thebibliography}




\appendix


\bsp	
\label{lastpage}
\end{document}